# Increasing city safety awareness regarding disruptive traffic stream

Olivera Kotevska

University of Grenoble Alpes, France

**Abstract:** Transportation systems serve the people in essence, in this study we focus in traffic information related to violation events to respond to safety requirements of the cities. Traffic violation events have an important role in city safety awareness and secure travel. In this work, we describe the use of knowledge discovery from traffic violation reports in combination with demographics approach using inductive logic programming to automatically extract knowledge about traffic violation behavior and their impact on the environment.

**1. Introduction**

Transportation systems serve the people in essence, in a modern intelligent transportation system it is significantly important to meet city and citizen's needs. Road traffic safety has significant impact of our daily live, because we all use different types of transportation, such as car, buses, or metro every day [1][3]. Even when the road conditions are perfect, some traffic violations could happen. Road traffic safety deals with a complexity of various factors and combination of them, that can have influence on it, such as infrastructure, distractive driving, traffic intense, climate conditions [4][5][6]. Quantifying local areas based on traffic data is intrinsically difficult due to the problem of assigning the mobile traffic incidents to locations.

This work is inspired by the need for the development of methods for complex event detection and processing in urban environment. The main goal is to investigate methods for extracting the useful information from heterogeneous data and find the rules for processing and predictive complex events in traffic violations. We consider a variety of factors such as weather, demographics, spatial and temporal. If a variety of factors are considered there is a limit to the degree to which relationship between decreasing traffic violation events and those factors can be correlated using statistical analysis. This study therefore used inductive models developed using artificial intelligence (AI) based techniques for traffic violation reasoning. *Inductive reasoning* is a logical process in which multiple premises, all believed true or found true most of the time, are combined to obtain a specific conclusion. We apply Inductive Logic Programming (ILP) [2] to the data because ILP can more flexibly learn rules than other machine-learning methods. ILP can easily and logically express the relationships among complex features. It can derive rules in forms that we can easily understand. Decision Trees can also derive rules in forms that can easily understand, but ILP has the advantages that it can learn rules flexibility using background knowledge represented by predicate logic, and that it can discover rules from a multi relational database consisting of multiple tables.

Therefore, the point of this study was to extract traffic rules for identifying the traffic violation events from time-series traffic violation data using ILP. We focus on Montgomery County, Maryland, because by the Farmers Insurance Group three cities from Montgomery County (MC), Maryland (MD) are considered as the most secure large metropolitan areas for 2013, they are Bethesda, Gaithersburg, and Frederick[1]. Since Bethesda and Gaithersburg belong to MC, MD we focus on them to identify the rules for the identifying occurrence of traffic violation event. Traffic events unusual behavior are caused by series of other unpredictable events, they are usually extremely difficult to explain and harder to predict. Nevertheless, there are some cases in which it is obvious that one event can affect the other. Also, we have attempted to determine how weather and demographic information influence other complex events in traffic violations [7].

We visualize the data from four aspects such as spatial characterization of the events, over time dynamics, driver characteristics and consequences of the event in order to understand it better. We process the data in a form that ILP can understand, such as creating the rules based on the founding from the previous mentioned descriptive analysis and integration with other data sources such as demographics and weather. Then we apply ILP and as a result our ILP system has successfully extracted rules to decrease traffic violation events. Learned rules indicate that a combination of location context and demographic data is actually sufficient to identify occurrence of traffic violation. We believe that when we drive based on these learned rules, we can maintain satisfied safety level. ILP techniques have been used in the area of health care [8], save sensor navigation between indoor and outdoor system [9], and identifying relation between users in social networks [10]. In our latest knowledge this is the first time that this technique was used in the domain of traffic violations.

This paper is organized as follows; Section 2 describes areas on which this study is focused and problem definition. Section 3 presents descriptive event analytics for Montgomery County, Maryland, while section 4 gives more detailed view of the cities in MC. Chapter 5 concludes the study and presents future work.

## 2. Focal topics for this study

For this study are used data sets related to Montgomery County, Maryland, U.S.A. Main data set of interest is the traffic violations, while weather, and demographics were used to better understand the behavior of the traffic events. Figure 1 presents the relation between data sets entities.

*Figure 1. Relation between traffic violation event data entries*

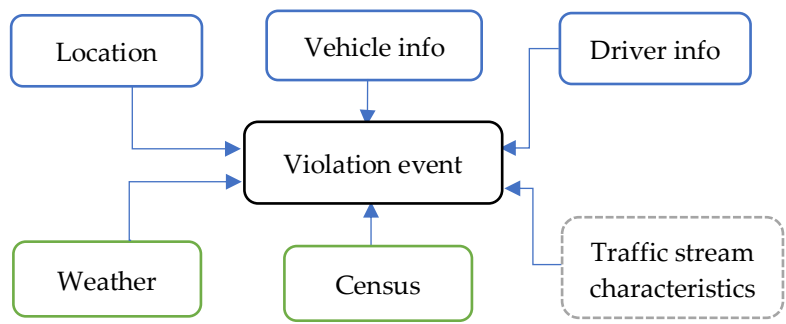

Traffic incidents data set[2]: We collected 190,117 records of traffic incidents reported throughout Montgomery County for the period of 1/1/2017 to 31/12/2017. The data is categorized in four logical groups such as Location which includes city name, longitude, latitude, work zone, residential, and green zone, Driver information such as personal injury, wearing a belt of not, Vehicle information such as commercial vehicle or not, and Traffic violation information such as type, charge, contributed to accident.

---

[1] https://www.farmers.com/news/2013/2013-most-secure-cities/
[2] https://data.montgomerycountymd.gov/Public-Safety/Traffic-Violations/4mse-ku6q

Weather dataset[3]: Daily data were collected over the same period as the crime dataset, for the cities in Montgomery County, Maryland (MD), U.S.A. Montgomery County is covered by three weather centers (College Park Airport, MD, Ronald Reagan Washington National Airport, VA and Montgomery County Airpark, MD). For this analysis, data from the College Park Airport, MD was selected because it covers most of the cities under investigation.

Census dataset: Census data was collected for cities in Montgomery County, Maryland, U.S.A. including demographics properties such as population count, education degree bachelor or higher, and median household income.

**3. Traffic deviation events of interest**

Traffic stream is characterized with three variables speed, density, and flow. Normal flow is affected by many factors such as number of lanes, intersections present along the road, percentage of heavy vehicles, and human factors[4]. Interrupted flow can happen because of intersection signalized or unsignalized, due to signs or merging of two road or highways. But sometimes it is happening because of some other external interruptions that have direct influence on the traffic flow such as police man is pulling over a car because of high speed, distracted driving, and accidents.

Traffic violations are the most common types of offense that people make. They are typically divided into two types: major and minor. Both types have consequences, depending on the severity of the violation[5]. Violation types can be Citation, Warning, and ESERO (Electronic Safety Equipment Repair Order). The most occurred traffic violation events in Montgomery County (MC), Maryland (MD) are presented in Table 1 and half of them are Citation type.

*Table 1. Most occurred traffic violations events in Montgomery County, Maryland.*

| Traffic violation event description | Total number of events |
|---|---|
| 1. Driver failure to obey properly placed traffic control device instructions | 16057 |
| 2. Failure to display registration card upon demand by police officer | 8779 |
| 3. Driver using hands to use handheld telephone while motor vehicle is in motion | 5904 |
| 4. Displaying expired registration plate issued by any state | 5004 |
| 5. Failure of individual driving on highway to display license to uniformed police on demand | 4957 |
| 6. Driving vehicle on highway with suspended registration | 4405 |
| 7. Driver failure to stop at stop sign line | 3986 |
| 8. Failure to obey stop light signal | 3473 |
| 9. Driving vehicle on highway without current registration plates and validation tabs | 3602 |
| 10. Exceeding the posted speed limit of 40 mph | 3323 |

We are going to focus on understanding these most occurred violation events and have a detailed analysis to some of the reasons for this distractive driving in order to better understand the events of interest and identify the rules of event occurrence. These events can be categorized in two categories such as events

---

[3] www.wunderground.com
[4] U.S. Department of Transportation, Federal Highway Administration: https://www.fhwa.dot.gov/publications/research/operations/tft/
[5] Maryland MVA Point System and Penalties (Last visit 11/28/2018) https://www.courts.state.md.us/sites/default/files/import/district/forms/criminal/dccr090.pdf

where the driver was not respecting the traffic road rules such as events 1, 3, 7, 8, and 10 is category 1, and category 2 driver not having the right equipment and documents such as events 2, 4, 5, 6, and 9.

*3.1. Distractive driving*

Distracted driving is any activity that derives driver attention from driving. Based on Maryland Department of Transportation Motor Vehicle Administration (MVA)[6] there are four types of distractions visual, auditory, manual, and cognitive. Texting while driving is especially dangerous because it includes three of the types of distractions [5] and based on National Highway Traffic Safety Administration [6] sending or reading messages takes of your eyes from the road for about 5 seconds long enough to pass football field 55 mph. While using hand-held phone while driving is prohibited in most of the states in US[7], some drivers think they are making a safe choice by using a hands-free device. This approach in fact distract the brain and potentially includes two of the types of distractions[8]. Based on National Highway Traffic Safety Administration in 2016 37,461 lives[9] were lost and 3,450 were because of distractive driving, while 391,000 were injured in 2015, and 58% of them are of teens[10]. Texting while driving or "Driver using hands to use handheld telephone while motor vehicle is in motion" is one of the ten most occurred traffic violations in the cities in MC, MD. We analyze the events in the category 1 in few aspects in order to better understand them and possibly determine a factor that influence them such as (a) consequences of the traffic violation event, (b) driver state characteristics, (c) temporal characteristics, (d) spatial characteristics, and (e) vehicle characterization.

a) Consequences of distractive driving are split in three groups, from them contributed to accident were 210, personal injury 108, and 138 cause a property damage. Table 2 has more detailed information. From these events in 40 cases driver was using a phone while driving.

*Table 2. Traffic violation events consequences.*

| Property damage | Contributed to accident | Personal injury | Total |
|---|---|---|---|
| No | No | Yes | 54 |
| No | Yes | No | 93 |
| No | Yes | Yes | 54 |
| Yes | No | No | 75 |
| Yes | Yes | No | 63 |

b) Driver state characteristics in these events are: in 3 cases of them the driver was under the influence of alcohol and in 937 cases was not wearing a belt. Also, 38% or 12,381 of the drivers were female and 62% or 20,359 were male. From the race aspect white race is dominant with 12,355 (36%) participants, black are 9,120 (27%), Hispanic 6,651 (20%), Asian 2,443 (<1%), Native American 74, and Other are 2,100 (<1%).

---

[6] Maryland Department of Transportation Motor Vehicle Administration (Last visit: 11/28/2018) http://www.mva.maryland.gov/safety/distracteddriving.htm
[7] American Automobile Association (Last visit: 11/28/2018) https://drivinglaws.aaa.com/tag/distracted-driving/
[8] National Safety Council (Last visit: 11/28/2018) https://www.nsc.org/road-safety/safety-topics/distracted-driving/cell-phone-distracted-driving
[9] National Highway Traffic Safety Administration (Last visit: 11/28/2018) https://www.nhtsa.gov/press-releases/usdot-releases-2016-fatal-traffic-crash-data
[10] American Automobile Association (Last visit: 11/28/2018) https://newsroom.aaa.com/2015/03/distraction-teen-crashes-even-worse-thought/

c) Temporal analysis of traffic violations is represented by interpretations in hours, weekdays, and months per year as factors that potentially plays role. Figure 2 shows the events dynamic based on hour and weekdays.

*Figure 2. Interpreting traffic violations based on hour and weekdays.*

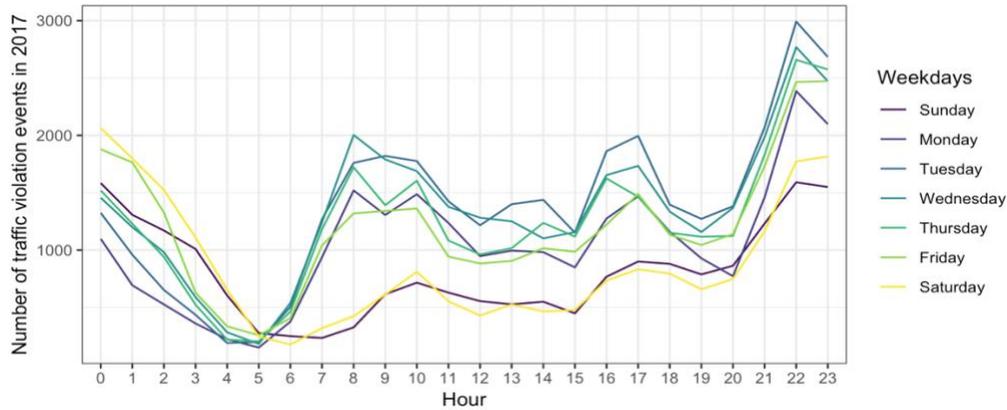

From figure 2 we can notice that during the day there are three picks in the number of traffic violations. One in the morning and late afternoon or 7-9 am and 16-17 pm probably due to the rush hours and another bigger pick in the evening around 10 pm. And from figure 3 we can notice that this trend continues over the whole year.

*Figure 3. Interpreting number of traffic violations per month with days in a week and hours.*

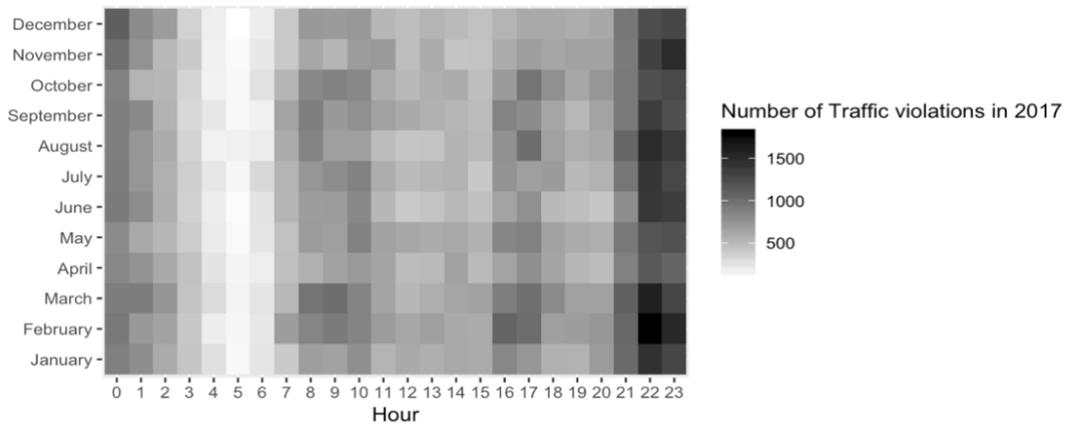

From the figure 2 we can observe that number of events in working days is higher than weekends and Friday has less traffic events compared with other working days. While, during the winter (November, December, January) and summer season (June, July, and August) number of events is decreased compared to spring (February, March, April, and May) and fall season (September, October).

d) Spatial analysis is interpreted by visualizing the areas where events happened. We mapped the locations that have more then 10 events occurred at the same location.

*Figure 4. Interpreting geo-location (longitude and latitude) density of traffic violation events.*

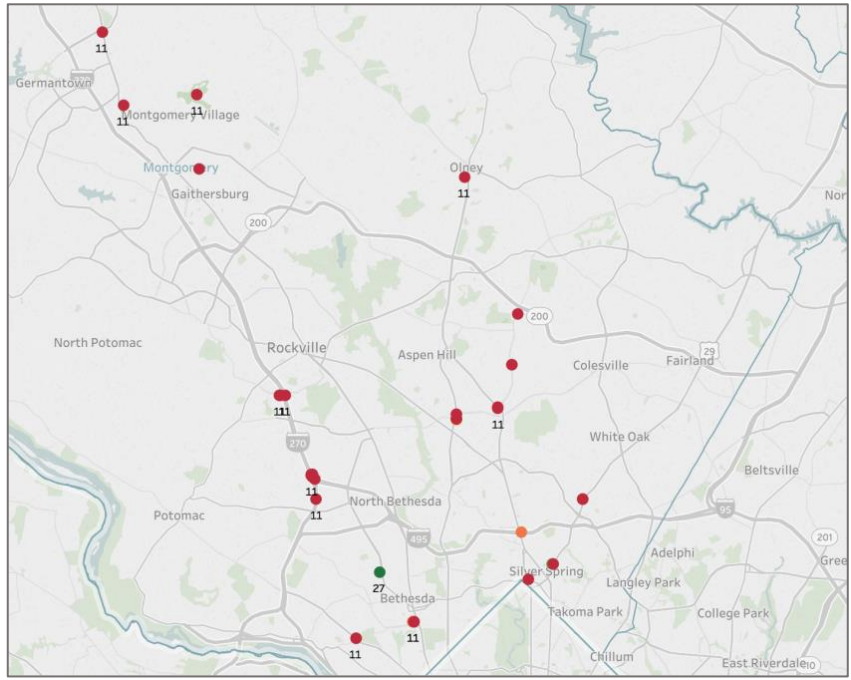

When we looked closely into the most frequent locations they are on the main road or close to the main roads and intersections, close to the green areas, or close to community areas (people visit frequently). One of the main roads are Connecticut Avenue and Wisconsin Avenue and community places such as hospitals and park, see figure 5.

*Figure 5. Context representation of the most affected locations.*

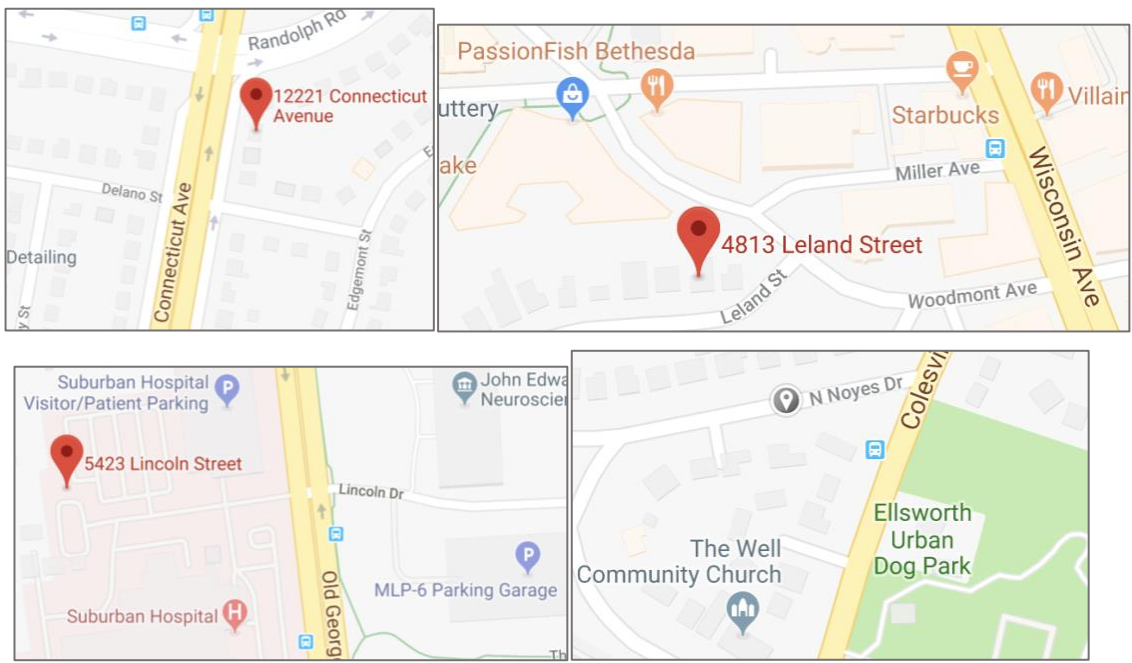

The location where mostly drivers do not follow the role of not using the phone while driving is in the location close to athletic center and the location where usually they exceed the speed is close to the county club and intersection and in the same location close to athletic center, see figure 6.

*Figure 6. Most affected location with the event of using a phone while driving.*

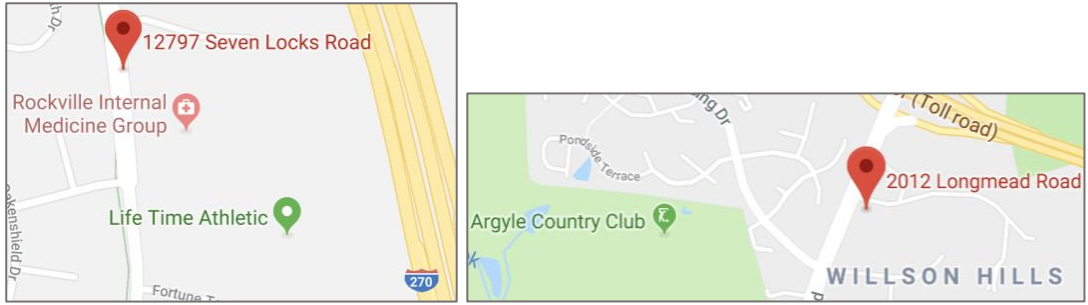

e) Vechicle characteristics involved in the traffic violation events are mostly automobiles but there are significat number of motorcycle and trucks involved, see table 3.

*Table 3. Number and type of vehicle types involved in traffic violation event. "Other" category represents the rest of the vehicle types involved in traffic violation event dataset.*

| Vehicle Type | Automobile | Light Duty Truck | Other | Station Wagon | Motorcycle | Heavy Duty Truck | Recreational Vechicle |
|---|---|---|---|---|---|---|---|
| Category 1 | 29,147 | 1,856 | 627 | 381 | 228 | 199 | 184 |

From them 210 contributed to accidents and 194 of the cases were by automobile, 5 of them are by light duty truck and 1 by heavy duty truck, 3 times by motorcycle, and 2 times by bus. The detailed view is automobile characteristics are type of model usually Toyota and Honda in 36% of the cases with black or silver color. Among the trucks usually are Ford. In general, usual models are Toyota, Honda, Ford, and Nissan are involved with more than 1000 events.

Number of events when driver was using a phone while driving is 5,904 and most of them were by automobile, see table 4. In 23 cases the driver contributed to accident mostly by the automobile.

*Table 4. Type of vehicles involved in traffic violation event - Driver using hands to use handheld telephone while motor vehicle is in motion.*

| Vehicle type | Number of events | Vehicle type | Number of events |
|---|---|---|---|
| Automobile | 5031 | Recreational Vehicle | 43 |
| Light Trucks | 509 | Heavy Duty Truck | 31 |
| Other | 159 | Motorcycle | 21 |
| Station Wagon | 103 | Tractor | 7 |

*3.2. Discussion*

Traffic violation events are serious health problems; to reduce them it would be helpful to quickly identify any regions and activities that have potential to become risk factor to public safety. We found that spatial component is decision factor, analysis of the roads at the most affect locations in terms of the number of lanes shows that they are usually at the one or two-line one direction lines but the environmental context is more important factor such community areas. Night hours is another factor we used for the analysis, data shows that there is a correlation between events occurred during night hours and increase of the night time except for June and July, see figure 7.

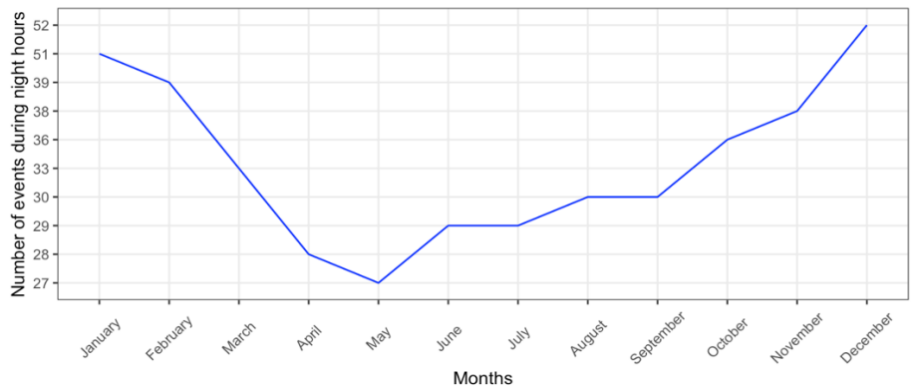

*Figure 7. Number of events during the night hours per month in 2017.*

Weather analysis show that there is a trend over the whole year during that at certain hours number of traffic events drops or increase. Since this trend continues during the whole year even when the weather is changing from summer to winter. When we separate the data per season we noticed that there is increase of disturbed driving events during the February and March and another time in August and September. While the biggest decrease is in December and July. So, the weather is not a relevant factor but maybe this behavior is related to the school year and break.

From the demographic point of view Maryland state have a population of 6.1 million people. Montgomery County is the most populated county with 1 million residents in 19 cities, towns, and villages and number of traffic violation events in 2017 compared with the population is 14.3%.

**4. Characterization of cities in several dimensions**

Micro and Macro trends of road traffic safety aspects are to view big data from a slightly different angle, we are interested in the ability to zoom between the micro level of analysis (an individual object) and the macro level (a collection) to see what new knowledge allows you to expose, and the stories it lets you tell. Previously we show the characteristic of Montgomery County from the perspective of traffic violations, now we would go at the micro or city level characterization. We focus on two cities since they are recognized as the safest cities to live Bethesda and Gaithersburg. Cities are characterized by the demographics, properties such as land area, water area, number of schools and hospitals, traffic events properties are number and type of traffic events, age of the population, most occurred traffic violations, vehicles involved, spatial and temporal aspects, see table 5 for more details.

*Table 5. Comparison between two cities - Bethesda and Gaithersburg.*

| | Comparison properties | Bethesda | Gaithersburg |
|---|---|---|---|
| Traffic events properties | Traffic violations | 2432 (3%) | 3688 (6%) |
| | Most occurred traffic events | Failure to obey instructions | Failure to obey instructions |
| | Second most occurred traffic event | Using phone while driving | Exceeding speed |
| | Night hours statistical parameters | Mean = 4.9 | Mean = 6.5 |
| | Light duty truck involved | 6.5% | 9% |
| | Cars old less then 10 years | 62.3% | 54.5% |
| | Location statistical parameters | Mean = 2.6 | Mean = 3.8 |
| Demographics | Population | 60858 | 59933 |
| | Density | 1624 km$^2$ | 2571 km$^2$ |
| | Education-higher then high school | 83.7% | 53.3% |
| | Median household income | 154.559 | 85.773 |
| | Poverty | 2.8% | 9.5% |
| | Younger then 65 and older then 18 | 64.8% | 58.3% |
| Driver | Personal injury | 13 | 10 |
| | Contributed to accident | 21 | 16 |
| | Belt | 6 | 290 |
| City properties | Area land | 13.1 km$^2$ | 26.72 km$^2$ |
| | Area water | 0.1 km$^2$ | 0.3 km$^2$ |
| | Schools | 18 | 25 |
| | Hospitals | 3 | 0 |
| | Main road | I-495 | I-270 |

Although the population number is very close for both cities the number of traffic violation events is higher in Gaithersburg. The comparison analysis based on some of the demographic factors shows that there is a difference in education and median household income properties between them and the correlation is high. While the most occurred event is the same for both of the cities there is a difference in the second event in Bethesda it is "using phone while driving" and in Gaithersburg is "exceeding speed". While from the temporal perspective number of events that happen during the daylight hours is higher in Bethesda then Gaithersburg. Statistical parameters for Bethesda are mean is 4.9, median 5.5, variance 6.4, and standard deviation is 2.5. While for Gaithersburg are mean is 6.5, median 6.5, variance 13, and standard deviation is 3.6. But when we look into the location context usually in Bethesda is in the intersections and restaurant/shopping areas while in Gaithersburg most of the locations are in restaurant/shopping areas. The distribution of the number of events per location is much more disperse in Gaithersburg then Bethesda, see figure 8 and 9. In Bethesda there are many events that occurred only once per location. Statistical parameters for the event distribution for Bethesda are mean is 2.6, variance 25, and standard deviation 5, while for Gaithersburg are mean 3.8, variance 122, and standard deviation 11.

*Figure 8. Number of events per location in Gaithersburg.*

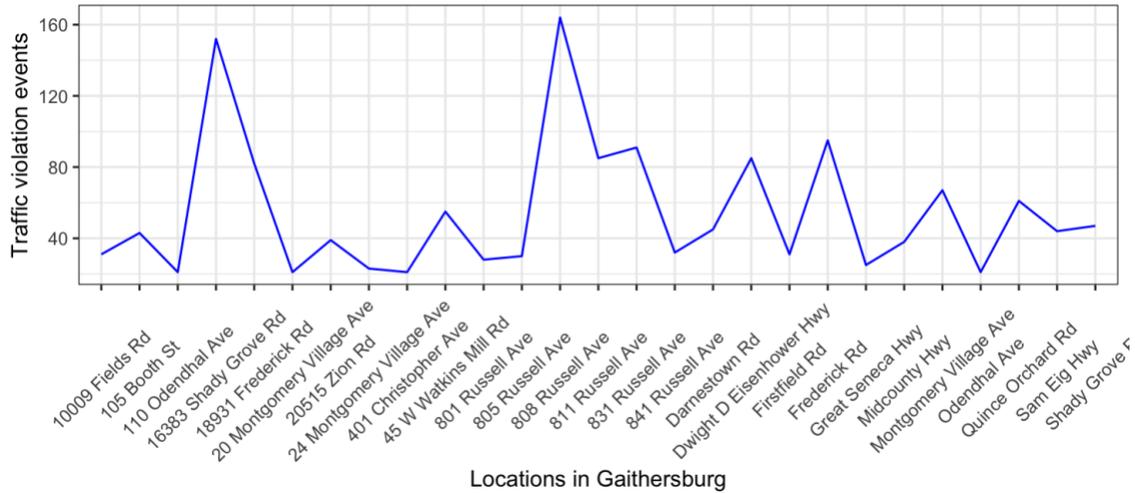

*Figure 9. Number of events per location in Bethesda.*

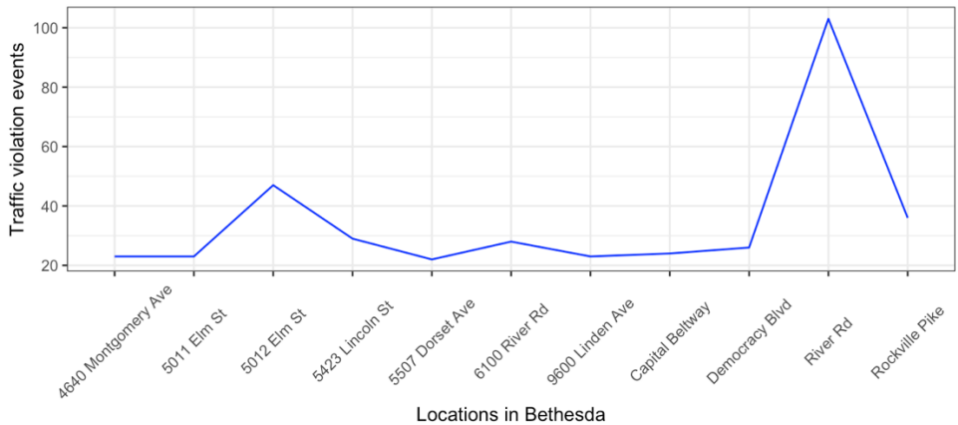

Based on this we develop the background knowledge relations for the ILP analysis. We identified the following potential hypothesis. Those rules are:

*Table 6. Construction of background knowledge and training examples.*

| Training examples | Background knowledge |
| --- | --- |
| night_hours(Bethesda, 4.9, 5.5, 6.4, 2.5) | main_road("exceeding speed", I-495) |
| night_hours(Gaithersburg, 6.5, 6.5, 13, 3.6) | main_road("exceeding speed", I-270) |
| location_distribution(Bethesda, 2.6, 25, 5) | population_density(Bethesda, 1624) |
| location_distribution(Gaithersburg, 3.8, 122, 11) | population_density(Gaithersburg, 2571) |
| location_context(Bethesda, intersection) | median_income(Bethesda, >150000) |
| location_context(Gaithersburg, community areas) | median_income(Gaithersburg, 75000-150000) |
| event_previous_occurence("using phone", >20) | education(Bethesda, >80%) |
| e_previous_occurence("exceeding speed", >20) | education(Gaithersburg, 50%-80%) |
| driver_characteristics(belt="yes") | |
| vehicle_year(>2009) | |

Rule 1: *is_event_inBethesda(X) :- event_time(X, 8 pm),*
*event_period_of_year('November'),*
*location_context('Athletic center'),*
*event_previous_occurence(X, >10),*
*vehicle_year(>2009),*
*driver_characteristics(belt='yes').*

The rule states that if finding X was at 8 pm on November close to the athletic center, there were prior events on the same location more then 10, and driver was wearing a belt then probably the event happened in Bethesda city.

Rule 2: *is_event_inGaithersburg(X) :- main_road(X, I-270),*
*event_previous_occurence(X, >20),*
*driver_characteristics(belt='no').*

The rule states that if finding X was close to highway and there were prior events on the same location more then 20, and driver was not wearing a belt then probably the event happened in Gaithersburg city.

Rule 3: *safe_location(Y, Bethesda) :- event_previous_occurence(Y, <5),*
*location_context(Bethesda, Y, community areas),*
*event_type (X, Y),*
*night_hours(Y, ).*

Rule 4: *event_happen(X, Y):- education(Y, >80%),*
*median_income(Y, >150000),*
*poverty(Y, <3%),*
*density(Y, <2000 km$^2$),*
*past_event_probability(X, ~3%).*

Rules 4 is generalization rule if we know the demographics about the city such as education, median income, poverty, density, and event probability in the cities with similar properties then we can expect the number of events to be similar.

Using traffic violation events to describe event occurrence and safer locations difficult as the number of traffic violation events is a function of multiple attributes and values. Certain type of traffic events is more predictable such as "Driver failure to obey properly placed traffic control device instruction". But because of the nature some of the violation events are hard to predict. We do not have information about the traffic flow density and the age of the drivers which can be an important factor.

## 5. Conclusions and future work

This study presents an application of ILP in the fields of public safety, namely, characterization of traffic violation effect on the environment. The knowledge discovered by ILP should be helpful for the design of further research experiments in safer neighborhood, and awareness about different traffic violations. Better pattern detection to better plan routes, schedules and so forth. In addition, we have shown that the ILP approach can be effectively used for detecting traffic violation rules.

These rules can improve mobility in the regions and build a more sustainable transportation network. Considering rush hours, traffic violations and location context could help the city better deploy resources and funding towards areas that are important to its citizens. For some of the event's local representatives possibly can-do changes so they will not happen in the future, while for some of the events such as exceeding speed limit we can create a service to inform the people for safety issues, to be more careful or to avoid.

We are interested to see if this behavior and rules can be applied to other cities in United States of America (U.S.A.) with similar properties as well. And if it is possible to make generalization with other states in U.S.A. and to characterize the other counties in the same state.

In the future, we plan to enhance the background knowledge by including more detailed information, as well as additional data. By integrating richer and more relevant background knowledge, we hope to not only improve the classification of traffic violation rules from open data police reports, but also to shed light on the complex relationships that exist between traffic violation and crime in order to improve the city safety and behavior in urban areas.

**Acknowledgments:** Greatest acknowledgment for Dr. Jack Boudreaux for his feedback, discussions, review, and time.